\documentclass[12pt]{article}
\usepackage[dvips]{graphicx}
\usepackage{pdproc}

\newcommand{\mweak}{M_{\text{weak}}}
 
\newcommand{\mstar}{M_{*}} 

\newcommand{\OmegaDM}{\Omega_{\text{DM}}}
\newcommand{\ifb}{\text{fb}^{-1}} 

\newcommand{\kev}{\text{keV}} 

\newcommand{\gev}{\text{GeV}} 
\newcommand{\tev}{\text{TeV}}

\newcommand{\m}{\text{m}}

\newcommand{\s}{\text{s}}
\newcommand{\yr}{\text{yr}}

\newcommand{\etal}{{\em et al.}}

\newcommand{\eqref}[1]{Eq.~(\ref{#1})}

\newcommand{\figref}[1]{Fig.~\ref{fig:#1}}

\newcommand{\NLSP}{\text{NLSP}}

\newcommand{\mNLSP}{m_{\NLSP}}

\newcommand{\mgravitino}{m_{\gravitino}}

\newcommand{\gravitino}{\tilde{G}}

\newcommand{\mgaugino}{M_{1/2}}

\def\agt{\mathrel{\raise.3ex\hbox{$>$\kern-.75em\lower1ex\hbox{$\sim$}}}}
\def\alt{\mathrel{\raise.3ex\hbox{$<$\kern-.75em\lower1ex\hbox{$\sim$}}}}
\newcommand{\text}[1]{{\rm #1}}
\newcommand{\mtext}[1]{\mbox{{\rm #1}}}

  \makeatletter 
  \def\@cite#1{[#1]} 
  \makeatother    
\begin{document}

\renewcommand{\thefootnote}{\alph{footnote}}


\title{ SuperWIMP Cosmology and Collider Physics\footnote{Plenary talk
given by JLF at SUSY2004, the 12th International Conference on
Supersymmetry and Unification of Fundamental Interactions, Tsukuba,
Japan, 17-23 June 2004.}  }

\author{JONATHAN L. FENG$^*$, ARVIND RAJARAMAN$^*$, BRYAN T. SMITH$^*$, \\
SHUFANG SU$^\dagger$, FUMIHIRO TAKAYAMA$^*$ }

\address{ 
$^*$Department of Physics and Astronomy, University of California \\
Irvine, CA 92697, USA \\
$^\dagger$Department of Physics, University of Arizona \\
Tucson, AZ 85721, USA
}

\abstract{Dark matter may be composed of superWIMPs,
superweakly-interacting massive particles produced in the late decays
of other particles.  We focus here on the well-motivated
supersymmetric example of gravitino LSPs.  Gravitino superWIMPs share
several virtues with the well-known case of neutralino dark matter:
they are present in the same supersymmetric frameworks (supergravity
with $R$-parity conservation) and naturally have the desired relic
density.  In contrast to neutralinos, however, gravitino superWIMPs
are impossible to detect by conventional dark matter searches, may
explain an existing discrepancy in Big Bang nucleosynthesis, predict
observable distortions in the cosmic microwave background, and imply
spectacular signals at future particle colliders. }

\normalsize\baselineskip=15pt

\section{Introduction}
\label{sec:introduction}

In recent years, there has been tremendous progress in understanding
the universe on the largest scales.  In particular, the energy density
in non-baryonic dark matter is known to be~\cite{Spergel:2003cb}
\begin{equation}
\OmegaDM = 0.23 \pm 0.04 
\end{equation}
in units of the critical density.  At the same time, we have no idea
what the microscopic identity of non-baryonic dark matter is.  The
dark matter problem therefore provides precise, unambiguous evidence
for new physics and has motivated new particles such as
axions~\cite{Peccei:1977ur,Wilczek:pj,Weinberg:1977ma},
neutralinos~\cite{Goldberg:1983nd,Ellis:1983ew}, Q
balls~\cite{Kusenko:1997si}, wimpzillas~\cite{Chung:1998ua},
axinos~\cite{Covi:1999ty}, self-interacting dark
matter~\cite{Spergel:1999mh}, annihilating dark
matter~\cite{Kaplinghat:2000vt}, Kaluza-Klein dark
matter~\cite{Servant:2002aq,Cheng:2002ej},
branons~\cite{Cembranos:2003mr,Cembranos:2003fu}, and many others.

Here we review a new class of dark matter candidates: superWIMPs,
superweakly-interacting massive particles produced in the late decays
of other
particles~\cite{Feng:2003xh,Feng:2003uy,Feng:2003nr}. SuperWIMPs have
several strong motivations:
\begin{itemize}
\item They are present in well-motivated frameworks for new physics,
  including models with supersymmetry (supergravity with $R$-parity
  conservation) and extra dimensions (universal extra dimensions with
  KK-parity conservation).
\item Their relic density is naturally in the right range to be dark
  matter without the need to introduce and fine-tune new energy
  scales.
\item They can explain an existing anomaly, namely the observed
  underabundance of $^7$Li relative to the prediction of standard Big
  Bang nucleosynthesis.
\item They have rich implications for early universe cosmology and
  imply spectacular signals at the Large Hadron Collider (LHC) and the
  International Linear Collider (ILC).
\end{itemize}

\section{The Basic Idea}
\label{sec:basic}

As noted above, superWIMPs exist in theories with supersymmetry and in
models with extra dimensions.  We concentrate on the supersymmetric
scenarios here.  Details of the extra dimensional realizations may be
found in Refs.~\cite{Feng:2003xh,Feng:2003uy,Feng:2003nr}.

In the simplest supersymmetric models, supersymmetry is transmitted to
standard model superpartners through gravitational interactions, and
supersymmetry is broken at a high scale.  The mass of the gravitino
$\gravitino$ is
\begin{equation}
m_{\gravitino} = \frac{F}{\sqrt{3} \mstar} \ ,
\label{gravitinomass}
\end{equation}
and the masses of standard model superpartners are
\begin{equation}
\tilde{m} \sim \frac{F}{\mstar} \ ,
\label{tildem}
\end{equation}
where $\mstar = (8 \pi G_N)^{-1/2} \simeq 2.4 \times 10^{18}~\gev$ is
the reduced Planck scale and $F \sim (10^{11}~\gev)^2$ is the
supersymmetry breaking scale squared.  The precise ordering of masses
depends on unknown, presumably ${\cal O}(1)$, constants in
\eqref{tildem}.  Most supergravity studies assume that the lightest
supersymmetric particle (LSP) is a standard model superpartner, such
as the neutralino.  In this case, it is well-known that the neutralino
naturally freezes out with a relic density that is in the right range
to account for dark matter~\cite{reviews}.

The gravitino may be the LSP, however~\cite{Feng:2003xh,Feng:2003uy,%
Feng:2003nr,Ellis:2003dn,Buchmuller:2004rq,Wang:2004ib,Feng:2004zu,%
Feng:2004mt,Feng:2004gn,Ellis:2004bx,Roszkowski:2004jd,Bi:2003qa}.  In
supergravity, the gravitino has weak scale mass $\mweak \sim 100~\gev$
and couplings suppressed by $\mstar$.  The gravitino's extremely weak
interactions imply that it is irrelevant during thermal freeze out.
The next-to-lightest supersymmetric particle (NLSP) therefore freezes
out as usual, and if the NLSP is a slepton, sneutrino, or neutralino,
its thermal relic density is again $\Omega_{\NLSP} \sim 0.1$.
However, eventually the NLSP decays to its standard model partner and
the gravitino.  The resulting gravitino relic density is
\begin{equation}
\Omega_{\gravitino} = \frac{m_{\gravitino}}{\mNLSP} \Omega_{\NLSP} \ .
\end{equation}
In supergravity, where $m_{\gravitino} \sim \mNLSP$, the gravitino
therefore inherits a relic density of the right order to be much or
all of non-baryonic dark matter.  The superWIMP gravitino scenario
preserves the prime virtue of WIMPs, namely that they give the desired
amount of dark matter without relying on the introduction of new,
fine-tuned energy scales.\footnote{In this aspect, the superWIMP
scenario differs markedly from previous gravitino dark matter
scenarios.  Gravitinos are the original supersymmetric dark matter
candidates~\cite{Pagels:ke,Weinberg:zq,Krauss:1983ik,%
Nanopoulos:1983up,Khlopov:pf,Ellis:1984eq,Ellis:1984er,%
Juszkiewicz:gg,Ellis:1990nb,Moroi:1993mb,Bolz:2000fu,Khlopov:rs}.
Previously, however, gravitinos were expected to be produced either
thermally, with $\Omega_{\gravitino} \sim 0.1$ obtained by requiring
$\mgravitino \sim \kev$, or through reheating, with
$\Omega_{\gravitino} \sim 0.1$ obtained by tuning the reheat
temperature to $T_{\text{RH}} \sim 10^{10}~\gev$.}

Because superWIMP gravitinos interact only gravitationally, with
couplings suppressed by $\mstar$, they are impossible to detect in
conventional direct and indirect dark matter search experiments.  At
the same time, the extraordinarily weak couplings of superWIMPs imply
other testable signals.  The NLSP is a weak-scale particle decaying
gravitationally and so has a natural lifetime of
\begin{equation}
\frac{\mstar^2}{\mweak^3} \sim 10^4 - 10^8~\s \ .
\label{lifetime}
\end{equation}
This decay time, outlandishly long by particle physics standards,
implies testable cosmological signals, as well as novel signatures at
colliders.

\section{Cosmology}
\label{sec:cosmology}

The most sensitive probes of late decays with lifetimes in the range
given in \eqref{lifetime} are from Big Bang nucleosynthesis (BBN) and
the Planckian spectrum of the cosmic microwave background (CMB).  The
impact of late decays to gravitinos on BBN and the CMB are determined
by only two parameters: the lifetime of NLSP decays and the energy
released in these decays.  The energy released is quickly thermalized,
and so the cosmological signals are insensitive to the details of the
energy spectrum and are determined essentially only by the total
energy released.

The width for the decay of a slepton to a gravitino is
\begin{equation}
 \Gamma(\tilde{l} \to l \tilde{G}) =\frac{1}{48\pi \mstar^2}
 \frac{m_{\tilde{l}}^5}{m_{\tilde{G}}^2} 
 \left[1 -\frac{m_{\tilde{G}}^2}{m_{\tilde{l}}^2} \right]^4 \ ,
\label{sfermionwidth}
\end{equation}
assuming the lepton mass is negligible. (Similar expressions hold for
the decays of a neutralino NLSP.)  This decay width depends on only
the slepton mass, the gravitino mass, and the Planck mass.  In many
supersymmetric decays, dynamics brings a dependence on many
supersymmetry parameters.  In contrast, as decays to the gravitino are
gravitational, dynamics is determined by masses, and so no additional
parameters enter.  In particular, there is no dependence on left-right
mixing or flavor mixing in the slepton sector.  For $\mgravitino /
m_{\tilde{l}} \approx 1$, the slepton decay lifetime is
\begin{eqnarray}
 \tau(\tilde{l} \to l \tilde{G})
\simeq 3.6\times 10^8~\s
\left[\frac{100~\gev}{m_{\tilde{l}} - m_{\gravitino}}\right]^4
\left[\frac{m_{\tilde{G}}}{\tev}\right]\ .
\label{eq:decaylifetime}
\end{eqnarray}
This expression is valid only when the gravitino and slepton are
nearly degenerate, but it is a useful guide and verifies the rough
estimate of \eqref{lifetime}.

The energy release is conveniently expressed in terms of
\begin{eqnarray}
\xi_{\text{EM}} \equiv \epsilon_{\text{EM}} B_{\text{EM}}
Y_{\text{NLSP}}
\label{eq:xi_EM}
\end{eqnarray}
for electromagnetic energy, with a similar expression for hadronic
energy.  Here $\epsilon_{\text{EM}}$ is the initial EM energy released
in NLSP decay, and $B_{\text{EM}}$ is the branching fraction of NLSP
decay into EM components.  $Y_{\text{NLSP}} \equiv
n_{\text{NLSP}}/n_{\gamma}$ is the NLSP number density just before
NLSP decay, normalized to the background photon number density
$n_{\gamma} = 2 \zeta(3) T^3 / \pi^2$.  It can be expressed in terms
of the superWIMP abundance:
\begin{equation}
Y_{\text{NLSP}}\simeq 3.0 \times 10^{-12}
\left[\frac{\tev}{m_{\gravitino}}\right]
\left[\frac{\Omega_{\gravitino}}{0.23}\right] \ .
\label{eq:def_Y}
\end{equation}

Once an NLSP candidate is specified, and assuming superWIMPs make up
all of the dark matter, with $\Omega_{\gravitino} = \Omega_{\text{DM}}
= 0.23$, the early universe signals are completely determined by only
two parameters: $m_{\gravitino}$ and $m_{\NLSP}$.

\subsection{BBN Electromagnetic Constraints}

BBN predicts primordial light element abundances in terms of one free
parameter, the baryon-to-photon ratio $\eta \equiv n_B / n_{\gamma}$.
In the past, the fact that the observed D, $^4$He, $^3$He, and $^7$Li
abundances could be accommodated by a single choice of $\eta$ was a
well-known triumph of standard Big Bang cosmology.

More recently, BBN baryometry has been supplemented by CMB data, which
alone yields $\eta_{10} = \eta / 10^{-10} = 6.1 \pm
0.4$~\cite{Spergel:2003cb}.  This value agrees precisely with the
value of $\eta$ determined by D, considered by many to be the most
reliable BBN baryometer.  However, it highlights slight
inconsistencies in the BBN data.  Most striking is the case of $^7$Li.
For $\eta_{10} = 6.0\pm 0.5$, the value favored by the combined D and
CMB observations, the standard BBN prediction is~\cite{Burles:2000zk}
\begin{eqnarray}
^7\text{Li/H} &=& 4.7_{-0.8}^{+0.9} \times 10^{-10} \label{Li}
\end{eqnarray}
at 95\% CL.  This contrasts with observations.
Three independent studies find 
\begin{eqnarray}
\mtext{$^7$Li/H} &=& 1.5_{-0.5}^{+0.9} \times 10^{-10} \quad 
\mtext{(95\% CL)~\cite{Thorburn}} \\
\mtext{$^7$Li/H} &=& 1.72_{-0.22}^{+0.28} \times 10^{-10} \ 
\mtext{($1\sigma + \text{sys}$)~\cite{Bonafacio}} \\
\mtext{$^7$Li/H} &=& 1.23_{-0.32}^{+0.68} \times 10^{-10} \ 
\mtext{(stat + sys, 95\% CL)~\cite{Ryan:1999vr}} \ ,
\end{eqnarray}
where depletion effects have been estimated and included in the last
value.  Within the published uncertainties, the observations are
consistent with each other but inconsistent with the theoretical
prediction of \eqref{Li}, with central values lower than predicted by
a factor of 3 to 4.  $^7$Li may be depleted from its primordial value
by astrophysical effects, for example, by rotational mixing in stars
that brings Lithium to the core where it may be
burned~\cite{Pinsonneault:1998nf,Vauclair:1998it}, but it is
controversial whether this effect is large enough to reconcile
observations with the BBN prediction~\cite{Ryan:1999vr}.

\begin{figure}[tb]
\begin{center}
\includegraphics*[width=11cm]{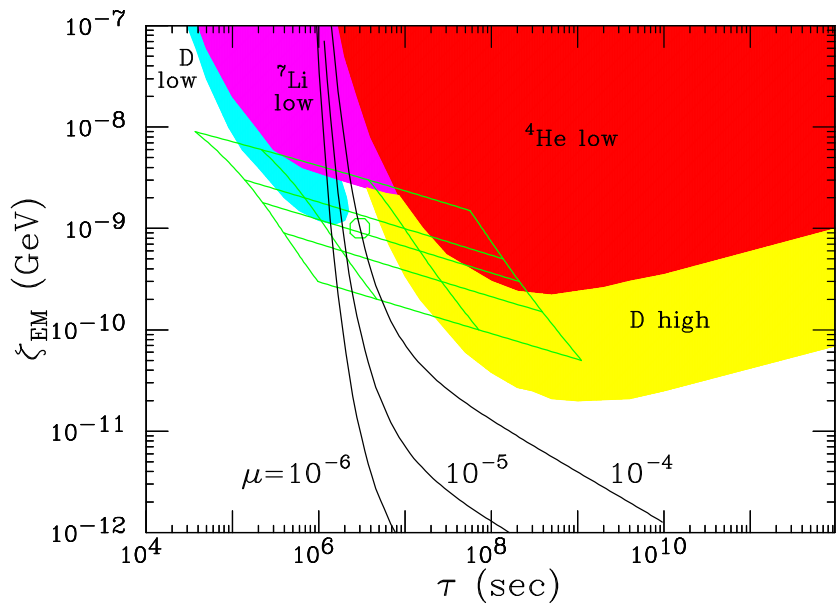}
\caption{Predicted and excluded regions of the $(\tau,
\zeta_{\text{EM}})$ plane in the superWIMP dark matter scenario, where
$\tau$ is the lifetime for $\tilde{l} \to l \tilde{G}$, and
$\zeta_{\text{EM}}$ is the normalized electromagnetic energy release.
The grid gives predicted values for $m_{\tilde{G}} = 100~\gev -
3~\tev$ (top to bottom) and $\Delta m \equiv m_{\tilde{l}} -
m_{\tilde{G}} = 600~\gev - 100~\gev$ (left to right), assuming
$\Omega_{\tilde{G}} = 0.23$.  BBN constraints exclude the shaded
regions; the circle indicates the best fit region where $^7$Li is
reduced to observed levels without upsetting other light element
abundances.  Contours of CMB $\mu$ distortions indicate the current
bound ($\mu < 0.9 \times 10^{-4}$) and the expected future sensitivity
of DIMES ($\mu \sim 10^{-6}$).  {}From
Ref.~\protect\cite{Feng:2003uy}.
}
\label{fig:stau}
\end{center}
\end{figure}

We now consider the effects of NLSP decays to gravitinos.  For WIMP
NLSPs, that is, sleptons, sneutrinos, and neutralinos, the energy
released is dominantly deposited in electromagnetic cascades.  For the
decay times of \eqref{lifetime}, mesons decay before they interact
hadronically.  The impact of EM energy on the light element abundances
has been studied in Refs.~\cite{Kawasaki:1994sc,Holtmann:1998gd,%
Kawasaki:2000qr,Cyburt:2002uv}.  The results of
Ref.~\cite{Cyburt:2002uv} are given in \figref{stau}.  The shaded
regions are excluded because they distort the light element abundances
too much.  The predictions of the superWIMP scenario for a stau NLSP
with $m_{\gravitino}$ and $\m_{\NLSP}$ varying over weak scale
parameters are given in \figref{stau} by the grid.

We find that the BBN constraint excludes some weak scale parameters.
However, much of the weak scale parameter space remains viable.  Note
also that, given the $^7$Li discrepancy, the best fit is not achieved
at $\xi_{\text{EM}} = 0$, but rather for $\tau \sim 3\times 10^6~\s$
and $\xi_{\text{EM}} \sim 10^{-9}~\gev$, where
$^7$Li is destroyed by late decays without
changing the other relic abundances.  This point is marked by the
circle in \figref{stau}.  The energy release predicted in the
superWIMP scenario naturally includes this region.  The $^7$Li anomaly
is naturally resolved in the superWIMP scenario by a stau NLSP with
$m_{\NLSP} \sim 700~\gev$ and $m_{\gravitino} \sim 500~\gev$.

\subsection{BBN Hadronic Constraints}

Hadronic energy release is also constrained by BBN~\cite{Reno:1987qw,%
Dimopoulos:1987fz,Dimopoulos:1988ue,Kohri:2001jx,Jedamzik:2004er,%
Kawasaki:2004yh,Kawasaki:2004qu}.  In fact, constraints on hadronic
energy release are so severe that even subdominant contributions to
hadronic energy may provide stringent constraints.

Slepton and sneutrino decays contribute to hadronic energy through the
higher order processes
\begin{eqnarray}
\tilde{l} &\to& l Z \tilde{G} \ , \ \nu W \tilde{G} \nonumber \\
\tilde{\nu} &\to& \nu Z \tilde{G} \ , \ l W \tilde{G} \ ,
\end{eqnarray}
when the $Z$ or $W$ decays hadronically.  These three-body decays may
be kinematically suppressed when $m_{\tilde{l}, \tilde{\nu}} -
m_{\tilde{G}} < m_W, m_Z$, but even in this case, four-body decays,
such as $\tilde{l} \to l \gamma^* \tilde{G} \to l q\bar{q} \tilde{G}$,
contribute to hadronic cascades and may be important.  The branching
fractions for these decays have been calculated in
Refs.~\cite{Feng:2004zu,Feng:2004mt}.  The end result is that these
constraints are stringent and important, as they exclude regions of
parameter space that would otherwise be allowed.  At the same time,
much of the parameter space in the case of slepton and sneutrino NLSPs
remains viable.  For details, see
Refs.~\cite{Feng:2004zu,Feng:2004mt,Shufang}.

In contrast to the case of slepton and sneutrino NLSPs, the neutralino
NLSP possibility is very severely constrained by bounds on hadronic
energy release.  This is because neutralinos contribute to hadronic
energy even through two-body decays 
\begin{equation}
\chi \to Z \gravitino , \ h \gravitino \ ,
\end{equation}
followed by $Z,h \to q \bar{q}$.  The resulting hadronic cascades
destroy BBN successes, and exclude this scenario unless such decays
are highly suppressed.  Kinematic suppression is not viable, however
--- if $m_{\chi} - m_{\gravitino} < m_Z$, the decay $\chi \to \gamma
\gravitino$ takes place so late that it violates bounds on EM
cascades.  Neutralino NLSPs are therefore allowed only when the
two-body decays to $Z$ and $h$ bosons are suppressed dynamically, as
when the neutralino is photino-like, a possibility that is not
well-motivated by high energy frameworks.

\subsection{CMB Constraints}

The injection of electromagnetic energy may also distort the frequency
dependence of the CMB black body radiation.  For the decay times of
interest, with redshifts $z \sim 10^5$ to $10^7$, the resulting
photons interact efficiently through $\gamma e^- \to \gamma e^-$ and
$e X \to e X \gamma$, where $X$ is an ion, but photon number is
conserved, since double Compton scattering $\gamma e^- \to \gamma
\gamma e^-$ is inefficient.  The spectrum therefore relaxes to
statistical but not thermodynamic equilibrium, resulting in a
Bose-Einstein distribution function
\begin{equation}
f_{\gamma}(E) = \frac{1}{e^{E/(kT) + \mu} - 1} \ ,
\end{equation}
with chemical potential $\mu \ne 0$.

In \figref{stau} we show contours of chemical potential $\mu$, as
determined by updating the analysis of Ref.~\cite{Hu:gc}.  The current
bound is $\mu < 9\times
10^{-5}$~\cite{Fixsen:1996nj,Eidelman:2004wy}. We see that, although
there are at present no indications of deviations from black body,
current limits are already sensitive to the superWIMP scenario, and
are even beginning to probe regions favored by the BBN considerations
described above. In the future, the Diffuse Microwave Emission Survey
(DIMES) may improve sensitivities to $\mu \approx 2 \times
10^{-6}$~\cite{DIMES}.  DIMES will therefore probe further into
superWIMP parameter space, and will effectively probe all of the
favored region where the $^7$Li underabundance is explained by decays
to superWIMPs.

\section{Colliders}
\label{sec:colliders}

As noted in \eqref{lifetime}, the next-to-lightest supersymmetric
particle (NLSP) decays to the gravitino with lifetime naturally in the
range $10^4 - 10^8~\s$.  However, as described above, cosmological
constraints exclude lifetimes at the upper end of this range and
disfavor neutralinos as NLSPs, leaving charged sleptons with lifetimes
below a year as the natural NLSP candidates.  The gravitino superWIMP
scenario therefore implies that the signal of supersymmetry at
colliders will be meta-stable sleptons with lifetimes of a month or a
year. This is a spectacular signal that will not escape notice at the
LHC~\cite{Drees:1990yw,Goity:1993ih,Nisati:1997gb,Feng:1997zr}.  In
addition, given the long lifetime, it suggests that decays to
gravitinos may be observed by trapping slepton NLSPs in water tanks
placed outside collider detectors and draining these tanks
periodically to underground reservoirs where slepton decays may be
observed in quiet environments.  

\begin{figure}[tb]
\begin{center}
\includegraphics*[width=11cm]{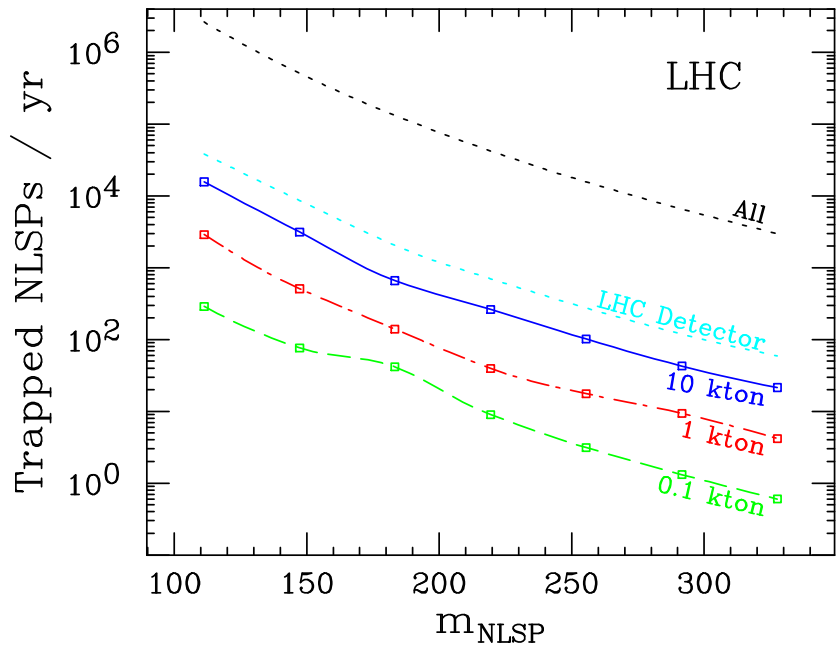}
\caption{The number of sleptons trapped per year at the LHC in water
tanks of size 10 kton (solid), 1 kton (dot-dashed), and 0.1 kton
(dashed).  The total number of sleptons produced is also shown (upper
dotted), along with the number of sleptons trapped in the LHC detector
(lower dotted).  These results assume luminosity $100~\ifb/~\yr$ and
minimal supergravity models with $\mgaugino = 300, 400, \ldots,
900~\gev$, $m_0 = 0$, $A_0 = 0$, $\tan\beta = 10$, and $\mu>0$. {}From
Ref.~\protect\cite{Feng:2004yi}.
\label{fig:LHCresults} }
\end{center}
\end{figure}

This possibility has been considered in
Refs.~\cite{Hamaguchi:2004df,Feng:2004yi}.  In
Ref.~\cite{Feng:2004yi}, we have explored the prospects for trapping
sleptons at the LHC by optimizing the water trap shape and placement
and considering a variety of sizes.  The results of Monte-Calo
simulations using the ISASUSY package~\cite{Paige:2003mg} are
displayed in \figref{LHCresults}.  The number that may be trapped is
highly model-dependent.  For minimal supergravity with $m_0= 0$, we
find that as many as $10^4$ staus may be stopped in a 10 kton trap
when the sleptons have mass around 100 GeV.  This is as light as is
allowed by current bounds.  For a less optimistic scenario with 219
GeV staus, hundreds and tens of sleptons may be caught each year in 10
kton and 1 kton traps, respectively.

The LHC results may be improved significantly if long-lived NLSP
sleptons are kinematically accessible at the ILC.  For the identical
case with 219 GeV sleptons discussed above, ${\cal O}(1000)$ sleptons
may be trapped each year in a 10 kton trap.  If only the NLSP is
accessible, this result may be achieved by tuning the beam energy so
that produced NLSPs barely escape the ILC detector.  The ability to
prepare initial states with well-known energies and the flexibility to
tune this energy are well-known advantages of the ILC.  Here, these
features are exploited in a qualitatively new way to produce slow
sleptons that are easily captured.

\begin{figure}[tbp]
\begin{center}
\includegraphics*[width=11cm]{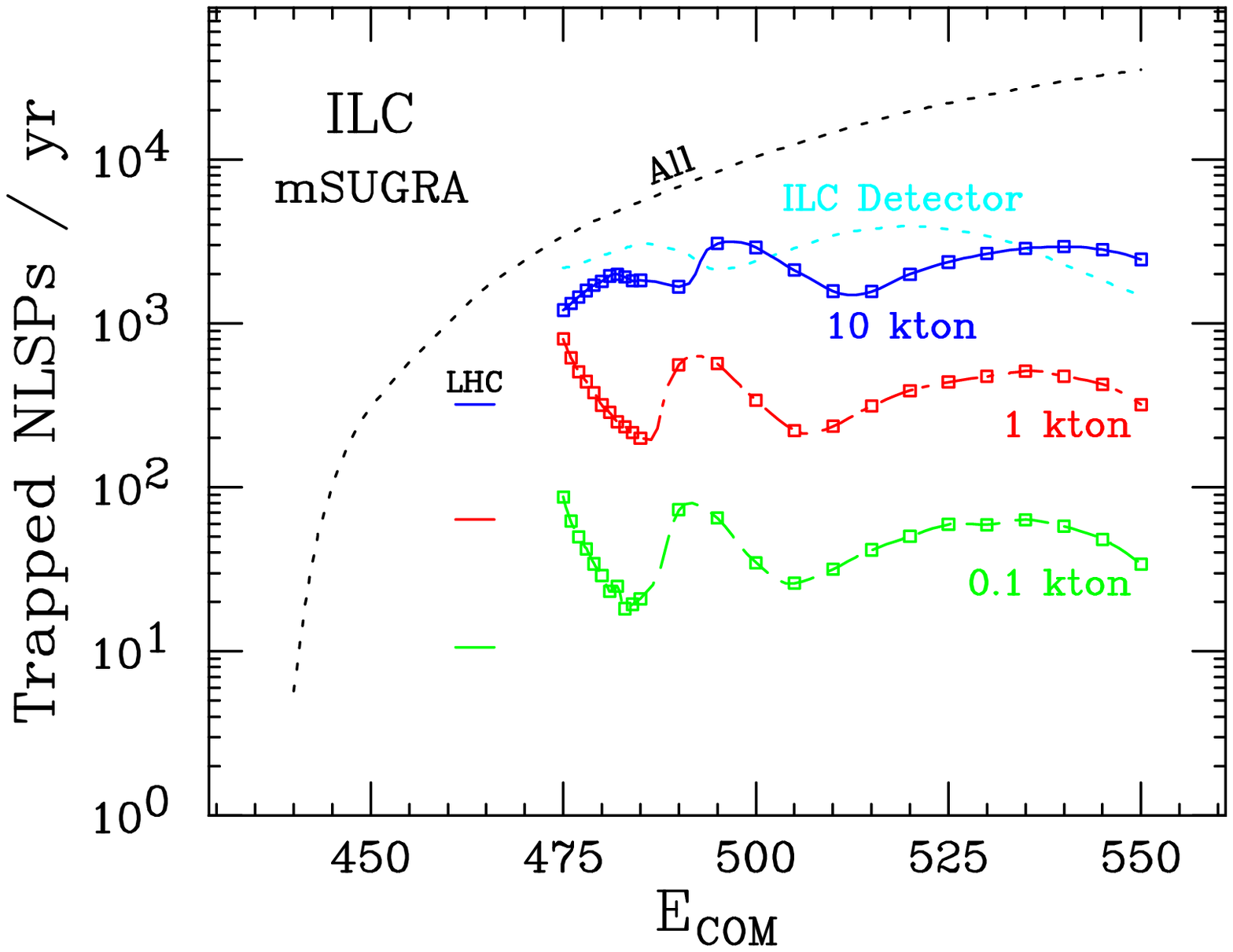}

\vspace*{.3in}

\includegraphics*[width=11cm]{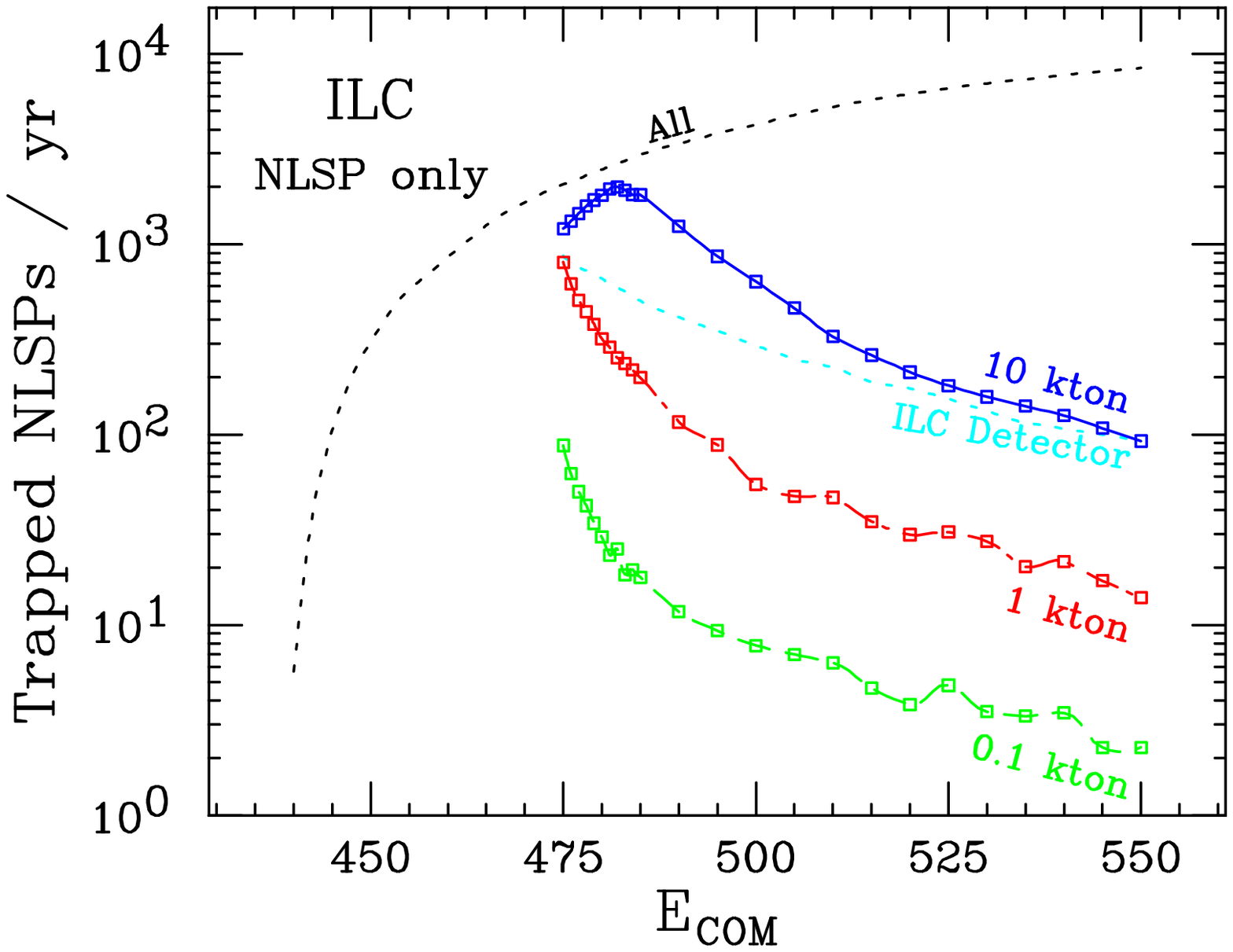}
\caption{The number of sleptons trapped per year at the ILC in 10 kton
(solid), 1 kton (dot-dashed), and 0.1 kton (dashed) water traps.  The
total number of sleptons produced is also shown (upper dotted) along
with the number of sleptons trapped in the ILC detector (lower
dotted).  The trap shape and placement have been optimized and we
assume luminosity $300~\ifb/\yr$.  In the top figure, the underlying
model is minimal supergravity with $\mgaugino = 600~\gev$, $m_0 = 0$,
$A_0 = 0$, $\tan\beta = 10$, and $\mu > 0$.  The LHC results for this
model are indicated.  In the bottom figure, the only accessible
superpartner is a 219 GeV NLSP stau. {}From
Ref.~\protect\cite{Feng:2004yi}.
\label{fig:ILCresults_NLSP} }
\end{center}
\end{figure}

If there are additional superpartner states accessible at the ILC,
even tuning the beam energy is not necessary.  The cascade decays of
other superpartner states produce a broad distribution of slepton
energies, and so for a broad range of beam energies, some sleptons
will be captured in the trap. We have noted also that, by considering
the slightly more general possibility of placing lead or other dense
material between the ILC detector and the slepton trap, an order of
magnitude enhancement may be possible, allowing up to ${\cal O}(10^4)$
sleptons to be trapped per ILC year.

The analysis here is valid with minor revisions for traps composed of
any material.  For concreteness we have considered traps composed of
water tanks, with the expectation that sleptons caught in water will
be easily concentrated and/or moved to quiet environments.  Other
possibilities, such as that considered in
Ref.~\cite{Hamaguchi:2004df}, are, however, well worth exploring.

\section{Conclusions}
\label{sec:conclusions}

Although the implications of supergravity for cosmology and particle
physics have been considered in great detail for decades, most work
has been centered on scenarios in which the LSP is a standard model
superpartner.  Here we have explored the gravitino LSP scenario.
Recent work has found significant cosmological motivations for this
possibility, as the gravitino may explain dark matter, and the
scenario may resolve current difficulties in Big Bang nucleosynthesis.

Despite the extremely weak couplings of superWIMPs, this scenario has
striking implications for cosmology and particle physics.  In
cosmology, the scenario predicts possible $\mu$ distortions in the CMB
spectrum at a level that will be probed by the planned DIMES mission.
Such distortions would provide significant corroborating evidence for
superWIMP dark matter.
   
The superWIMP scenario also implies long-lived charged particles,
which will provide a spectacular signal at colliders.  As discussed
above, such particles may be trapped and their decays studied in
detail.  There are many significant implications of such studies.  In
the case of supersymmetry and gravitino superWIMPs, these implications
have been considered in detail in
Refs.~\cite{Buchmuller:2004rq,Feng:2004gn}.  Briefly, simply by
counting the number of slepton decays as a function of time, the
slepton lifetime may be determined with high accuracy.  Given
thousands of sleptons, we expect a determination at the few percent
level.  The slepton decay width of \eqref{sfermionwidth} is a simple
function of the slepton and gravitino masses, and the slepton mass
will be constrained by analysis of the collider event kinematics. A
measurement of the slepton width therefore implies a high precision
measurement of the gravitino mass and, through \eqref{gravitinomass},
the supersymmetry breaking scale $F$.  Such measurements will provide
precision determinations of the relic density of superWIMP gravitino
dark matter, the contribution of supersymmetry breaking to vacuum
energy, and the opportunity for laboratory studies of late decay
phenomena relevant for Big Bang nucleosynthesis and the cosmic
microwave background.

The gravitino mass may also be determined, although not necessarily on
an event-by-event basis, by measuring the energy of slepton decay
products.  This provides a consistency check.  Alternatively, these
two methods, when combined, determine not only $m_{\gravitino}$, but
also the Planck mass $\mstar$.  This then provides a precision
measurement of Newton's constant on unprecedentedly small scales, and
the opportunity for a quantitative test of supergravity relations.

\section{Acknowledgements}

JLF thanks the organizers of SUSY04 for a beautifully organized and
stimulating conference. We are grateful to H.-C. Cheng and H.~Murayama
for valuable conversations, and thank A.~Lankford, W.~Molzon,
F.~Moortgat, D.~Stoker, and especially D.~Casper for experimental
insights.  The work of JLF was supported in part by National Science
Foundation CAREER Grant PHY--0239817, and in part by the Alfred
P.~Sloan Foundation.

\bibliographystyle{plain}

\end{document}